\documentstyle[aasms4,12pt]{article}

\begin{document}

\title{Star Forming Objects in the Tidal Tails of Compact Groups}
\author{J. Iglesias-P\'{a}ramo}
\author{jiglesia@ll.iac.es}
\affil{Instituto de Astrof\'{\i}sica de Canarias\\
38200 La Laguna, Tenerife, SPAIN}
\and
\author{J.M. V\'{\i}lchez}
\author{jvm@iaa.es}
\affil{Instituto de Astrof\'{\i}sica de Andaluc\'{\i}a (CSIC)\\
Apdo. 3004, 18080 Granada, SPAIN}
\vspace{.2in}

\begin{abstract}

A search for star forming objects belonging to tidal tails has
been carried out in a sample of deep H$\alpha$ images of 16 compact groups
of galaxies. A total of 36 objects with H$\alpha$ luminosity larger than
10$^{38}~erg~s^{-1}$ have been detected in five groups. 
The fraction of the total H$\alpha$ luminosity of their respective
parent galaxies shown by the tidal objects is always below 5\% except
for the tidal features of HCG~95, whose
H$\alpha$ luminosity amounts to 65\% of the total luminosity.
Out of this 36 objects, 9 star forming tidal dwarf galaxy
candidates have been finally identified on the basis of their projected distances
to the nuclei of the parent galaxies and their total H$\alpha$ luminosities.
Overall, the observed properties of the candidates resemble those 
previously reported
for the so-called tidal dwarf galaxies.

\end{abstract}


\section{Introduction}

Interactions between disk galaxies are known to produce strong
morphological disturbances and, if the geometry of the encounter is
favorable, to enhance the star formation rates of the disk
galaxies. The pioneering work by Tomre \& Tomre (1972) and a later study by Barnes
(1988) supported the
original idea by Zwicky (1956) that dwarf galaxies could originate
within the tidal tails resulting from a strong interaction between two
disk galaxies. Since then, several examples of dwarf galaxies with
tidal origin have already been detected in interacting systems in an
advanced stage of merging and also around galaxies
experiencing less dramatic interactions (Schweizer 1978, Mirabel et
al. 1991,1992, Hibbard \& van Gorkom 1996, Duc \& Mirabel 1997,1998).

The compact group environment has been suggested to be an ideal place for
interactions between galaxies given their large spatial densities and
their low velocity dispersions. Mendes de Oliveira \& Hickson (1994)
reported morphological signs of interaction in many galaxies of compact
groups from Hickson's Catalogue (1982). Other recent studies on the
photometric properties of the galaxies in Hickson Compact Groups at
different wavelengths resulted contradictory when compared to samples of
field galaxies: while CO, FIR, H$\alpha$ and optical fluxes seem to be
similar to those measured for field galaxies (Moles et al. 1994, Verdes-Montenegro et
al. 1998, Iglesias-P\'{a}ramo \& V\'{\i}lchez 1999),
radio continuum and 21cm emission are on
average lower than for samples of isolated galaxies (Williams \& Rood 1987,
Menon 1995, Huchtmeier 1997).

Hunsberger et al. (1998) carried
out a search for dwarf galaxy candidates in the close environment of a
sample of Hickson Compact Groups, by automated detection of faint
objects on deep $R$ band frames. After decontamination for the presence
of background galaxies, they found that the luminosity
function in compact groups showed an enhancement at the faint luminosity
end compared to a normal Schechter function. They concluded that the
initial dwarf galaxy population in compact groups is replenished by
``subsequent generations'' formed in the tidal debris of giant galaxy
interactions.

In this work we present the results of a search for tidal star forming
dwarfs in our H$\alpha$ images
of a sample of nearby compact groups (see V\'{\i}lchez \&
Iglesias-P\'{a}ramo 1998, Iglesias-P\'{a}ramo \& V\'{\i}lchez 1999). Five
of the groups were found to show optically prominent tidal tails with star forming objects
within them. We emphasize the fact that all the detected objects are
H$\alpha$ emitters, therefore implying that they are physically
associated with the galaxies of the group.
The main properties of the star forming objects will be presented in
\S\ref{sample}, as well as a description of their close
environment. \S\ref{discussion} contains a discussion on their
likely evolutionary scenario.

\section{The Sample\label{sample}}

Deep H$\alpha$ images for a total of 67 galaxies belonging to 16 Hickson
compact groups were inspected. 
Four telescopes were used in order to obtain all the data presented in
this paper: The INT 2.5m and JKT 1.0m in the Observatorio del Roque de los
Muchachos, the IAC80 in the Observatorio del Teide and the 2.2m
Telescope in the Observatorio Hispano-Alem\'{a}n de Calar Alto. Also,
some $R$ band continuum frames taken at
the 3.6m~CFHT were kindly made available to us by Dr. Paul Hickson. 
More details concerning the data acquisition
and reduction for the sample can be found in Iglesias-P\'{a}ramo \&
V\'{\i}lchez (1999)\footnote{Although not mentioned in that paper, HCG~80
belongs to the current sample}. 
H$\alpha$ emitting objects in tidal features were found in 5
groups of the sample: HCG~31, HCG~44, HCG~92, HCG~93 and HCG~95. We
note that although tidal features were reported for some other groups of
our sample 
(Mendes de Oliveira \& Hickson 1994), no H$\alpha$ objects have been
found associated with these features.

Figures~\ref{g31cont} to~\ref{g95cont} show detailed contour maps of
continuum subtracted H$\alpha$
images of the sections of the groups where star forming tidal objects were detected. The contour
scales are different for each group. 
A preselection of star forming objects was done 
by careful optical inspection of the H$\alpha$ images, with
the following criteria: (1) all the objects appear spatially resolved at the 
3$\sigma$ isophote over the
sky level, (2) they are more luminous than $10^{38}~erg~s^{-1}$ --
this somewhat arbitrary luminosity limit was chosen in
order to exclude normal low luminosity H{\sc ii} regions (see \S~3) --
(3) they are
located along tidal features, well away from the main body of the parent
galaxies. 

Table~\ref{prop} shows some physical properties of the tidal
objects. Column 1 shows the identifier of each object as quoted in
Figures 1 to 5 (with the
prefix TD added to the identifier); Column 2 shows
the name of the parent galaxy according to Hickson et al. (1989);
Column 3 shows the optical radius at the $25~mag~arcsec^{-2}$ isophote of the
parent galaxy, $R_{25}$; Column 4 shows
the measured H$\alpha$ luminosity in $erg~s^{-1}$; Column 5 shows
the fraction of the total H$\alpha$ luminosity emitted by the tidal
objects; Column 6 shows the H$\alpha$ equivalent width in \AA; Column 7
shows the projected distance of the tidal object to the center of the
parent galaxies normalized to $R_{25}$.
All the
distance-dependent quantities have been estimated adopting
$H_{0} = 75~km~s^{-1}~Mpc^{-1}$ and assuming a pure Hubble
flow\footnote{Note that the H$\alpha$ luminosities
published in our previous papers adopted a different value for
$H_{0}$}. Here follows a brief morphological description of the close
environment of the detected tidal objects.

\subsection{HCG~31}

This group, that hosts up to nine tidal objects, has been previously
discussed in the literature (i.e. Rubin et al. 1990, Iglesias-P\'{a}ramo
\& V\'{\i}lchez 1997a, Johnson et al. 1999). The parent galaxies seem to be in an
advanced stage of merging, and show strong star formation activity
everywhere. A third galaxy located to the West could be joining the interacting
pair. The whole group is immersed in an H{\sc i} cloud (Williams et
al. 1991) and is one of the few Hickson Compact Groups richer than
expected in H{\sc i}. The tidal tail containing objects TD~31c
to TD~31i, appears also delineated in the H{\sc i} map. The large equivalent
widths measured for some of the tidal objects, together with their
extremely blue broad band colors (Iglesias-P\'{a}ramo \& V\'{\i}lchez
1997a), suggest that we are probably observing two recently originated
dwarf galaxies. This is remarkable since this kind of objects were thought
to be born mostly in systems containing early-type galaxies (Duc \&
Mirabel 1998) without apparent optical spiral arms.

\subsection{HCG~44}

HCG~44 is quite a dispersed group. Overall, it was found to be deficient
in H{\sc i} (Williams \& Rood 1987). The individual galaxies HCG~44a and
c were also found to be H{\sc i} deficient. An H{\sc i} dwarf
galaxy was detected within the field of the group at approximately the
same radial velocity (Williams et al. 1991). The parent galaxy hosting the
tidal objects, HCG~44d, is a SB(s)c peculiar galaxy showing two distorted
tidal arms perpendicular to the direction of the central bar (Mendes de
Oliveira \& Hickson 1994). Almost no
underlying emission from the tidal objects was detected in broad band
images. The H{\sc i} map of this galaxy also
shows asymmetries and clear signatures of interactions (Williams et
al. 1991). The nearby galaxy HCG~44a, an early-type spiral with the plane
of the disk perpendicular to the tidal tails of HCG~44d, could be
responsible for the interaction suffered by HCG~44d; the
distortion detected in the outer disk of HCG~44a towards HCG~44d is
a relic of the interaction between both galaxies. 
Eight star forming objects were detected in the tidal tails of HCG~44d.

\subsection{HCG~92}

Several tidal objects were detected in this group, widely known as Stephan's
Quintet. Interesting reviews of the history and evolution of this
group can be found in Moles et al. (1997,1998) and Xu et al. (1999). 
The H{\sc i} distribution is decoupled from the optical galaxies
(Verdes-Montenegro et al., 2000), which could explain the low star formation
activity of this group in spite of the high rate of interaction among its
members. One of the tidal objects, TD~92a, is located at the tip of a long tidal
tail extending out from HCG~92c. This galaxy shows a Seyfert 2 type active nucleus,
a strong bar across the disk and several shells apparent in the
continuum image. However, excepting the nucleus,
no strong H$\alpha$ emission was detected throughout this galaxy. Xu
et al. (1999) performed a detailed study of the star forming regions in
HCG~92 based on IR imaging, and reported this tidal dwarf galaxy as
probably due to the interaction of HCG~92c with the nearby galaxy NGC~7320c,
not included as a member of the group in the original catalog. The rest
of the tidal objects were detected around the interacting pair of
galaxies HCG~92b and d. Each of these galaxies has developed one tidal
tail pointing towards the North. Our H$\alpha$ map of the tail
associated with HCG~92b shows a
vertical bar-like structure, probably containing shocked gas (Ohyama et
al. 1998) due to the intrusion of this galaxy into the group. At the tip
of this tail, a starburst region was reported by Xu et al. (1999) -- our
object TD~92b -- and attributed to the
collision of the galaxy with the intergalactic medium of the
group. We have detected several other objects in the outskirts of HCG~92d.

\subsection{HCG~93}

Only one galaxy, HCG~93b, hosts the seven tidal objects detected in this
group. HCG~93 is a very H{\sc i} poor compact group. The mean separation between its galaxies is
about 70~$kpc$, which makes it the largest of the five groups reported in this
work. The parent galaxy, HCG~93b, is an SB(s)cd peculiar galaxy with two
long tails, well visible in the optical continuum images. This galaxy,
together with the massive elliptical HCG~93a was cataloged as a binary
system by Peterson (1979). The low surface brightness asymmetric halo
surrounding HCG~93a (see the red continuum image in V\'{\i}lchez \&
Iglesias-P\'{a}ramo 1998) suggests that an interaction between the two
galaxies may be responsible for the tidal tails exhibited by HCG~93b as
well as for the faint halo of HCG~93a. Four of the tidal objects --
including the most luminous one -- were
detected along the Northern tail, and the remaining three along the
Southern one. 

\subsection{HCG~95}

In this group of galaxies, a three-galaxy interaction
seems to be taking place (Iglesias-P\'{a}ramo \& V\'{\i}lchez 1998). 
As a result, four tidal tails emerge from the
doubly nucleated galaxy HCG~95c. The gas distribution of the system
appears peculiar because it seems weighted towards the Eastern nucleus
of HCG~95c and the Northern tidal tail emerging from it, with almost no
gas in the rest of the system. Four tidal objects were detected along
the Northern gas-rich tail. A close inspection of the broad band and
H$\alpha$ images of the system suggested the possibility that transfer
of material was taking place from the most luminous object in this tidal
tail -- namely TD~95b -- towards the nucleus of the bright nearby
elliptical galaxy HCG~95a (Iglesias-P\'{a}ramo \& V\'{\i}lchez
1997b). The analysis of the broad band colors of the tidal objects
showed that the amount of underlying population present along
the Northern tidal tail probably decreases outwards, being almost negligible for
TD~95a and TD~95b.

\section{Discussion and Conclusions\label{discussion}}

As can be seen from Table~1, the H$\alpha$ equivalent widths and
luminosities of the
tidal objects suggest that most of them host very young star
formation bursts and are above the luminosity limit for giant H{\sc ii}
regions. 
The underlying continuum
luminosity varies greatly from object to object as shown in Figures~1 to
5. This
is the reason for the large range covered by their H$\alpha$ equivalent
width values, rather than the result of different evolutionary stages of the
corresponding star formation bursts of the tidal objects.
The fraction of the total H$\alpha$ flux corresponding to tidal objects is
typically under 5\% (see Table~1), except in the remarkable case of the tidal
features in HCG~95c, which amount to 65\% of the total
luminosity of the galaxy. We attribute this extraordinarily high
fraction of luminosity within the tidal tail to the fact that HCG~95c
belongs to a triple system showing a rather complicated interaction
sequence (Iglesias-P\'{a}ramo \& V\'{\i}lchez 1998).

The existence of these tidal objects is linked to
the possibility that they could subsequently evolve as independent dwarf galaxies, like
those observed at the tips of the tidal tails of the Antennae and the
Superantennae (Mirabel et al. 1991,1992). 
Looking to the optical images,
only rough morphological similarities between the tails of the
Antennae and Superantennae and the tails shown by our galaxies can be
found.
The Antennae and Superantennae fall in the most dramatic case
of interaction of a pair of galaxies, whereas our five groups span a
wide range of interaction strengths. 
At this point, two critical questions arise concerning the
possibility that tidal objects can evolve independently as dwarf galaxies: (1)
are they distant enough from the parent galaxy so as to be able to escape from the
potential well?, and (2) are they massive enough to be
self gravitating systems and to be considered dwarf galaxies?

Concerning the first question, some authors have claimed (Schweizer
1978; Hibbard \& van Gorkom 1996) that most
of the material ejected during an interaction is
accumulated in the tidal tails. 
N-body simulations suggest that a substantial fraction of
this material will slowly fall into the main body of the remnant of
the interaction, and only the outermost 20\% will probably gain
enough kinetic energy to be able to evolve
independently for a long time (see Hibbard \& Mihos 1995). 
Given that no information is available on the velocity
fields of the interacting pairs of our sample, and that
they show a great variety of interaction patterns, we will adopt a
quantitative criterion based on the projected distances of the tidal
objects from the nuclei of their
parent galaxies in order to select the most likely candidates to escape
from the potential well. Ferguson et al. (1998) carried
out a deep search for H{\sc ii} regions well outside the optical limits
of a sample of disk galaxies, and found no H{\sc ii} regions located at a
projected galactocentric distance larger than $2 \times R_{25}$. This
result was explained as due to the fact that the gas surface density is
below the critical density for star formation at such large radii from the
centers of the galaxies. Therefore, we can consider that those star
forming regions located that far from the nuclei of their parent galaxies
must have been originated by an external mechanism which has removed
gas from the inner regions of the parent galaxies. Star forming
regions far away from their parent galaxies may become kinematically
decoupled from the stellar disks; thus they are very likely escaping the potential
well of the parent galaxy and could become good tidal dwarf galaxy candidates.
According to
Table~\ref{prop}, {\em only 9 tidal objects fulfill this criterion} ($R > 2
\times R_{25}$).
Obviously, this is a conservative
criterion adopted given the lack of information on the velocity field of the
tidal tails. Therefore, we should bear in mind that because of
possible projection effects, the number of tidal dwarfs candidates
selected is a lower limit since other objects not satisfying this
criterion cannot be ruled out to be tidal dwarf galaxies. 

Concerning the second question posed above,
Elmegreen et al. (1993) proposed that
clouds as massive as $10^{8}~M_{\odot}$ can form during interactions of disk
galaxies. Under some conditions, these clouds could become
dwarf galaxies. In particular, when the mass of the perturber is
larger than approximately 1.4 times the mass  of the parent galaxy,
-- as appears to be the case for the five groups of our
sample presenting H$\alpha$ emitting tidal tails\footnote{estimated from a
comparison of the magnitudes of the parent galaxies with those of the
remaining galaxies in the groups} --
these models predict that the outermost clouds are very likely to escape the
potential well. 
Such $10^{8}~M_{\odot}$ clouds would produce $10^{6}~M_{\odot}$ starbursts,  
assuming a value for the efficiency of star formation of 1\% (see
Kennicutt 1998 and references therein). 
From the observational side, the H$\alpha$ luminosities reported for the
tidal dwarf galaxies of the Antennae (Mirabel et al. 1991) and NGC~5291 (Duc \& Mirabel
1998), are comparable to the ones derived in the present paper; the H{\sc i}
masses measured for those dwarfs appear to be typically above $10^{8}~M_{\odot}$.
In the same direction, from a large compilation of dwarf galaxies Hunter \& Gallagher (1986)
reported H{\sc i} masses larger than $10^{8}~M_{\odot}$  for
objects with $L_{\mbox{\scriptsize{H}}\alpha} \gtrsim
10^{39}~erg~s^{-1}$. Such bright objects are known to be
gravitationally bound, according to the well established
$L($H$\alpha$) {\em versus} $\sigma$ relationship for H{\sc ii}
galaxies (Terlevich \& Melnick 1981). The 9 objects selected above show
H$\alpha$ luminosities larger than $10^{39}~erg~s^{-1}$, which on
observational grounds seems to be a reasonable lower limit to ensure
self gravitation for H{\sc ii} complexes.

The two restrictions imposed above, though conservative, appear
reasonable since H{\sc ii} regions brighter than
$10^{39}~erg~s^{-1}$ are found to be located at galactocentric
distances lower than about $1.2 \times R_{25}$ (as derived from the
H{\sc ii} region surveys by
Rozas et al. 1996,2000).
A naive estimation of the dynamical time needed by our dwarf galaxy
candidates to travel a distance equivalent to $R_{25}$
as a consequence of
the interaction would be of the order of the fading time for H{\sc
ii} regions. Thus, we are confident that the 9 selected candidates must have
been generated during the interaction and do not belong to a
previous generation of disk H{\sc ii} regions.

From the arguments presented above, it appears clear that the 9 star forming
objects finally selected represent good candidates to satisfy the {\em escape} and
{\em self gravitation} conditions, and thus they can be included in the group of
the so-called tidal dwarf galaxies. Further information on total masses and relative
velocity with respect to their parent galaxies is required to confirm
that they satisfy both conditions.

It seems striking that the ratio between the number of candidate dwarf galaxies to the total
number of tidal objects detected, 25\%, appears consistent with the fraction of 
the ejected material which is expected to escape according to
theoretical models (Hibbard \& van Gorkom 1995). Since
all these candidate tidal galaxies were selected
at  projected distances larger than $2\times R_{25}$, we suggest that this last 
criterion might be a useful indicator of an {\em effective} escape
radius for these systems.

Summarizing, as a result of a search performed on a sample of net H$\alpha$ images of
16 compact groups of galaxies, we have identified 9 star forming
tidal objects proposed to be considered dwarf galaxy candidates given
their large projected distances from their parent galaxies and their large
H$\alpha$ luminosities.
These interesting objects deserve further detailed study, including dynamical and
evolutionary aspects which are beyond the scope of the present study.

\acknowledgements

We acknowledge John Beckman for carefully reading of the last
version of this document and for interesting comments and suggestions.
Thanks must also be given to A. Zurita for providing H{\sc ii} region catalogues.
The INT and the JKT are operated on the island of La Palma by the Isaac
Newton Group in
the Spanish Observatorio del Roque de Los Muchachos of the Instituto de
Astrof\'{\i}sica de Canarias.
The 2.2m Telescope is operated by the MPIA in
the Spanish Observatorio de Calar Alto.
This research has made use of the NASA/IPAC Extragalactic Database (NED)
which is operated by the Jet Propulsion Laboratory, California Institute
of Technology, under contract with the National Aeronautics and Space
Administration. This study was partly financed by the Spanish DGES
(Direcci\'{o}n General de Ense\~{n}anza Superior), grant PB97-0158.

\newpage






\figcaption[f1.eps]{Greyscale map of HCG~31 in the red continuum band. Overimpossed
contours correspond to the net H$\alpha$ emission. Tidal objects are
marked with arrows. Axis units correspond to arcseconds. North is
Up and East is Left.\label{g31cont}}

\figcaption[f2.eps]{Same as Figure~\ref{g31cont} for HCG~44.\label{g44cont}}

\figcaption[f3.eps]{Same as Figure~\ref{g31cont} for HCG~92.\label{g92cont}}

\figcaption[f4.eps]{Same as Figure~\ref{g31cont} for HCG~93.\label{g93cont}}

\figcaption[f5.eps]{Same as Figure~\ref{g31cont} for HCG~95.\label{g95cont}}

\newpage

\begin{table}
\begin{center}
\begin{tabular}{lcccccc}
\tableline
Name & Parent Gal. & $R_{25}$ & $L(\mbox{H}\alpha)$ & $x(\%)$ & $E.W.(\mbox{H}\alpha)$ &
$D_{proj}/R_{25}$ \\
 & & $kpc$ & $erg~s^{-1}$ & & \AA\ & \\
\tableline
TD~31a & HCG~31a, HCG~31c & 6.74 & 39.82 & 0.8 & 103 &  1.43 \\
TD~31b &      & & 39.96 & 1.1 &  50 &  1.03 \\
TD~31c &      & & 40.47 & 3.6 & 219 &  0.98 \\
TD~31d &      & & 39.21 & 0.2 &  -- &  1.37 \\
TD~31e &      & & 38.91 & 0.1 &  -- & 1.53 \\
TD~31f &      & & 38.91 & 0.1 &  -- & 1.69 \\
TD~31g &      & & 40.44 & 3.4 & 623 & 2.21 \\
TD~31h &      & & 40.06 & 1.4 & 592 & 2.52 \\
TD~31i &      & & 39.21 & 0.2 &  -- & 3.07 \\
\tableline
TD~44a & HCG~44d & 3.84 & 39.31 & 3.5 & 188 & 2.33 \\
TD~44b &      & & 39.35 & 3.8 & 163 & 1.59 \\
TD~44c &      & & 38.82 & 1.1 &  76 & 1.40 \\
TD~44d &      & & 38.68 & 0.8 & 105 & 1.14 \\
TD~44e &      & & 38.73 & 0.9 &  81 & 0.99 \\
TD~44f &      & & 38.25 & 0.3 & 118 & 0.97 \\
TD~44g &      & & 38.38 & 0.4 & 150 & 1.33 \\
TD~44h &      & & 38.08 & 0.2 &  -- & 1.22 \\
\tableline
TD~92a & HCG~92c & 19.51 & 40.36 & 2.8 &  90 & 2.41 \\
\tableline
TD~92b & HCG~92b, HCG~92d & 19.52 & 41.32 & 1.3 & 110 & 1.14 \\
TD~92c &      & & 41.08 & 0.7 &  95 & 1.64 \\
TD~92d &      & & 40.71 & 0.3 &  -- & 0.53 \\
TD~92e &      & & 41.11 & 0.8 &  30 & 0.36 \\
TD~92f &      & & 40.96 & 0.6 &  33 & 0.45 \\
TD~92g &      & & 40.74 & 0.3 &  -- & 1.07 \\
\tableline
TD~93a & HCG~93b & 10.90 & 39.06 & 0.3 & 23 & 3.02 \\
TD~93b &      & & 40.16 & 3.2 & 58 & 2.01 \\
TD~93c &      & & 39.39 & 0.6 & 12 & 1.78 \\
TD~93d &      & & 39.31 & 0.5 & 16 & 1.30 \\
TD~93e &      & & 40.75 & 5.0 & 22 & 0.94 \\
TD~93f &      & & 39.65 & 1.0 & 14 & 0.79 \\
TD~93g &      & & 39.48 & 0.7 & 12 & 1.07 \\
TD~93h &      & & 39.38 & 0.5 & 13 & 1.08 \\
\tableline
TD~95a & HCG~95c & 7.54 & 40.25 & 7.9 &  23 & 2.66 \\
TD~95b &       & & 40.91 & 36 & 179 & 2.02 \\
TD~95c &       & & 40.56 & 16 &  20 & 1.34 \\
TD~95d &       & & 40.20 & 7.2 &  14 &  0.51 \\
\tableline
\end{tabular}
\caption{Physical properties of the tidal dwarf objects: (1) Name of
the dwarfs; (2) Name of the parent galaxy according to Hickson et
al. (1989); (3) Radius in $kpc$ of the
$\mu_{B} = 25~mag~arcsec^{-2}$ isophote corresponding to the parent
galaxy; (4) H$\alpha$
luminosity in $erg~s^{-1}$; (5) Fraction of the total H$\alpha$
luminosity emitted by the dwarf galaxy; (6) H$\alpha$ equivalent width
of the dwarf galaxies, expressed in \AA; (7) Projected distance to the
center of the parent galaxy, in units of $R_{25}$.\label{prop}}
\end{center}
\end{table}

\end{document}